\documentclass[12pt]{iopart}
\usepackage{graphicx}
\usepackage{harvard}
\usepackage{enumerate}

\def\lapp{\ifmmode\stackrel{<}{_{\sim}}\else$\stackrel{<}{_{\sim}}$\fi}
\def\gapp{\ifmmode\stackrel{>}{_{\sim}}\else$\stackrel{>}{_{\sim}}$\fi}

\begin{document}

\title{The International Pulsar Timing Array}
\author{R. N. Manchester (for the IPTA)}
\address{CSIRO Astronomy and Space Science, PO Box 76, Epping NSW 1710, Australia}
\ead{dick.manchester@csiro.au}

\begin{abstract}
  The International Pulsar Timing Array (IPTA) is an organisation
  whose {\it raison d'\^etre} is to facilitate collaboration between
  the three main existing PTAs (the EPTA in Europe, NANOGrav in North
  America and the PPTA in Australia) in order to realise the benefits
  of combined PTA data sets in reaching the goals of PTA
  projects. Currently, shared data sets for 39 pulsars are available
  for IPTA-based projects. Operation of the IPTA is administered by a
  Steering Committee consisting of six members, two from each PTA,
  plus the immediate past Chair in a non-voting capacity. A
  Constitution and several Agreements define the framework for the
  collaboration. Web pages provide information both to members of
  participating PTAs and to the general public. With support from an
  NSF PIRE grant, the IPTA facilitates the organisation of annual
  Student Workshops and Science Meetings. These are very valuable both
  in training new students and in communicating current results from
  IPTA-based research.
 \end{abstract}

\maketitle

\section{Introduction}\label{sec:intro}
Pulsar timing arrays (PTAs) exploit the great period stability of
millisecond pulsars (MSPs) to explore a range of phenomena that
produce correlated timing variations among the pulsars in the
array. There are three main sources of correlated variations that
can be investigated: 
\begin{itemize}
\item Irregularities in reference time standards
\item Errors in the planetary ephemerides used for transferring
  observed pulse arrival times to the solar-system barycentre
\item Gravitational waves passing over the pulsars and the Earth
\end{itemize}
These three source categories have different spatial signatures, i.e.,
the correlated variations have different dependencies on the positions
of the pulsars. Consequently, to separate these different
source categories, a PTA must have a wide distribution of pulsars on
the sky (e.g., Foster \& Backer 1990).\nocite{fb90}

All pulse times of arrival (ToAs) are with reference to a time
standard, normally based on an international network of atomic
frequency standards. For example, ``Temps atomique international''
(TAI) is a uniform timescale, defined by the Bureau International des
Poids et Mesures (BIPM) in Paris, that is based on a weighted average of data
from several hundred atomic frequency standards around the
world. Normally time derived from a local observatory frequency
standard is referred to TAI using global navigation systems to effect
the time transfer. TAI has a stability of order 1:10$^{15}$ over
intervals of months to years \cite{app11}, comparable to the stability
of MSPs \cite{vbc+09}. Consequently it is feasible to use a PTA to
identify irregularities in TAI and similar atomic timescales. Such
irregularities will affect the period of all pulsars in the array in
the same way; if the reference timescale is running slow, all pulsars
will appear to be running fast and vice versa. In terms of a spatial
pattern, this is a monopole signature. 

To remove the effects of motion of the Earth from pulsar ToAs, all
pulsar timing analyses use a planetary ephemeris to calculate 
corresponding ToAs at the barycentre of the solar system,
assumed to be inertial (unaccelerated) with respect to the Local
Standard of Rest. Any error in the ephemeris results in an error in
the computed position of the Earth and hence residuals which are
positive for pulsars in the direction of the error and negative for
pulsars in the opposite direction. Such errors therefore have a dipole
signature on the sky. 

Because of the quadrupolar nature of gravitational radiation,
correlated residuals resulting from a gravitational
wave (GW) passing over the Earth have a different signature, positive for
pulsars separated on the sky by angles near zero and $\pi$ and
negative for separations near $\pi/2$ \cite{hd83}. 

\begin{figure}[ht]
\centerline{\includegraphics[width=120mm]{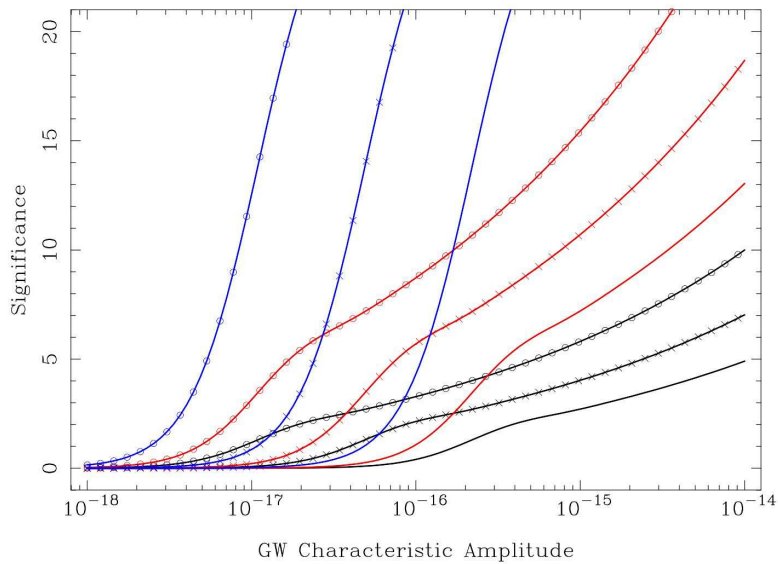}}
\caption{Significance of detecting an isotropic stochastic GW
  background using a PTA as a function of signal strength, number of
  pulsars in the PTA, and data spans. Black lines are for 20 pulsars,
  red lines 50 pulsars and blue lines 200 pulsars. Data spans are 5
  years (plain lines), 10 years (crosses) and 20 years (circles). Rms
  timining residuals of 100 ns and weekly sampling are
  assumed. \cite{mhb+13,vbc+09}}\label{fg:gw_sens}
\end{figure}

These different spatial signatures allow the different effects to be
separated. The significance of any detection depends in a complicated
way on the details of the signal, the distribution on the sky of
pulsars in the array, and the precision, cadence and span of the ToA
measurements. Figure~\ref{fg:gw_sens} illustrates this for the case of
detection of an isotropic stochastic GW background with a set of
idealised PTAs. While real PTAs may not reach
these ideal assumptions, the general trends remain valid.  Clearly the
most dramatic improvement in detection significance is obtained by
increasing the number of pulsars in the PTA, with sensitivity
approximately linearly proportional to the number of pulsars in the array
at a given GW amplitude. This is the core rationale for creation of
the International Pulsar Timing Array (IPTA).

There are three main PTAs currently operating: the European Pulsar
Timing Array (EPTA), the North American pulsar timing array (NANOGrav)
and the Parkes Pulsar Timing Array (PPTA) -- see articles in this
Focus issue by Michael Kramer \& David Champion, Maura McLaughlin and George Hobbs,
respectively. Each of these has been operating consistently since 2005
although earlier data from related pulsar timing programs are
frequently included in analyses. As is discussed in more detail in
\S\ref{sec:data} below, data are available for 39 pulsars, of which
11 are observed by two PTAs and eight by all three PTAs.

The benefits of combining data sets from the various PTAs have long
been recognised. However, the first concrete steps toward setting up a
framework to facilitate this were taken by Andrea Lommen and
colleagues when they organised the first IPTA meeting in Arecibo on
1-2 August, 2008. This meeting was very successful and resulted in a
draft Data Sharing Agreement between the three main PTAs. Informal
contact was maintained between representatives of the three PTAs over
the next couple of years and this led to the second IPTA Meeting, held
in Leiden in June 2010. This meeting, the first supported by the NSF
PIRE grant awarded to West Virginia University, Was in two parts, a student
workshop with pedagogical lectures and tutorial sessions, and a
science meeting with both invited and contributed talks and
posters. Following this meeting a representative IPTA Steering
Committee (IPTASC) was formed and this first met (by teleconference)
in February 2011. In its subsequent meetings, the IPTASC formalised
the goals and {\it modus operandi} of the IPTA as described in
\S\ref{sec:org} below.

One of the most important functions of the IPTA is to make available
to the wider community carefully calibrated combined data sets. As a
first step, data files have been made available by the three PTAs and
these are described in \S\ref{sec:data}. Another important function is
to coordinate and facilitate what have now become regular annual IPTA
student workshops and science meetings and to encourage outreach
programs related to IPTA objectives. These aspects of the IPTA are
discussed in \S\ref{sec:outreach}.

An early description of the IPTA was presented by George Hobbs at the
8th Edoardo Amaldi GW Conference held in New York in June 2009
\cite{haa+10}. The pulsars being observed by the three PTAs and the
expected ToA precisions were listed along with the expected GW
detection sensitivity for a combined IPTA data set.

\section{IPTA Organisation}\label{sec:org}
The IPTA in an organisation whose role is to facilitate cooperation
between the main existing PTAs in order to optimise progress toward
the mutual goals of PTA projects. It does this in several different
ways. Firstly it sets up projects with lead members from each of the
PTAs for specific goals related to use of combined data sets. One of
the most important of these projects has the goal of producing
combined data sets in a carefully calibrated and user-friendly
form. Secondly the IPTA facilitates the organisation of annual student
workshops and science meetings on PTA-related topics. Thirdly it
provides a framework for the establishment of working groups on topics
of interest to PTA researchers, including outreach programs and data
challenges. 

Operation of the IPTA is administered by a Steering Committee, the
IPTASC, which was formally established in February 2011. Each participating PTA
has two members on the Committee and the immediate past Chair has
ex-officio non-voting status. Table 1 gives the past and present
membership of the IPTASC. The IPTA meets approximately every two
months by teleconference or at the annual Science Meeting, and urgent
matters are discussed via email.

\begin{table}[ht]
{\scriptsize
\caption{Members of the IPTA Steering Committee}\label{tb:iptasc}
\begin{tabular}{lll} \hline
Year & \multicolumn{1}{c}{Chair} & \multicolumn{1}{c}{Members} \\ \hline
2011 & Andrea Lommen & Dick Manchester, Scott Ransom, Ben Stappers, Gilles Theureau,
Willem van Straten \\
2012 & Dick Manchester & Scott Ransom, Ingrid Stairs, Ben Stappers, Gilles Theureau,
Willem van Straten \\
2013 & Ben Stappers & Jim Cordes, Jason Hessels, George Hobbs, Scott
Ransom, Willem van Straten \\
\hline
\end{tabular}}
\end{table}

The main vehicle for communication of IPTA-related issues between PTA
researchers is a set of wiki-based web pages at
http://www.ipta4gw.org/wiki/. These are accessible only to registered
users who are normally members of a participating PTA, although a
``friend'' category with limited access exists. Currently there are 95
registered users. A new user can request access to the wiki from the
login page and such requests are normally promptly approved by the
IPTASC. The wiki has pages for policies and agreements, projects,
working groups, shared data, IPTASC Minutes and outreach. Registered
users have read and write access to most of these pages; friends have
this access only for the outreach page. IPTASC members and a few other
people have administration access to the wiki allowing them to modify
permissions, create new pages, approve user requests etc.

The IPTA operates under a {\bf Constitution} that was ratified by the IPTASC
in March 2012 and consists of the following items:

{\footnotesize
\begin{enumerate}[1]
\item The International Pulsar Timing Array (IPTA) is a consortium of
  existing Pulsar Timing Array (PTA) collaborations.
\item The aims of the IPTA are to facilitate collaboration between
  participating PTA groups and to promote progress toward PTA
  scientific goals.
\item The IPTA will host annual student workshops and science meetings
  open to both members and non-members.
\item An IPTA Steering Committee (IPTASC) consisting of one or two
  representatives of each participating PTA will establish policy
  guidelines for data sharing, publication of results based on shared
  data and other matters as the IPTASC decides.
\item New PTA collaborations may be admitted to the IPTA with the
  approval of the IPTASC. Depending on size, new collaborations may
  have either one or two representatives on the IPTASC.
\item Members of the IPTASC will normally serve two-year terms.
\item The Chair of the IPTASC will be elected by the IPTASC members,
  normally from within the IPTASC membership, and will serve a
  one-year term as Chair and a further one-year term as an ex-officio
  non-voting member.
\item The IPTASC will meet at regular intervals and at least three
  times per year to discuss and formulate IPTA policy.
\item Policy will normally be agreed by consensus within the IPTASC,
  but in the event of disagreement, a policy decision can be made with
  one dissension.
\item The IPTASC will establish policy documents on topics as agreed
  necessary.
\item The IPTASC will establish and maintain a list of IPTA
  projects. Each project will have a title, abstract and a list of
  lead authors for the associated publication. The Project list will
  be made available to all IPTA members through a password-protected
  website.
\item Projects can be proposed to the IPTASC at any time and will be
  reviewed at (or if urgent, before) the next IPTASC meeting.
\item The IPTASC will review the Project list at agreed intervals and,
  if deemed necessary, revise the goals or lead authorship of
  previously approved projects with not less than three months notice
  to the existing lead authors.
\item This Constitution may be amended by unanimous vote of the IPTASC
  not less than three months after notification of IPTA members of any
  proposed changes.
\end{enumerate}}

The next most important document governing the operation of the IPTA
is the {\bf Data Sharing Agreement}, ratified by the ad hoc IPTA Committee  in June
2009 following discussion at the Arecibo IPTA Meeting in August 2008,
which has the following terms and conditions:

{\footnotesize
\begin{enumerate}[1]
\item The data is made available to the PTAs for doing PTA work
  only. This could be GW work, ephemerides or new time standards, but
  it is not for studying the properties of individual sources.
\item Nothing is to be published from these data by people who didn't
  take the data without the participation of people from the group
  that did take the data, in both the analysis process and in any
  resulting publications.
\item IPTA-wide projects led by graduate students will be protected
  from prior publication by others. Such projects must be agreed to by
  the IPTA collaboration and the protection will be reviewed annually
  by the collaboration.
\item We share calibrated pulse profile data, timing template profiles
  and ToAs, but raw data files will be made available on request.
\item Shared data (including a summary table), lists of members of
  each PTA collaboration and lists of agreed collaborative projects
  will be made available on a server accessible only to members of the
  IPTA collaboration.
\item Data will be made available within six months of the date of
  observation.
\item Within a year of the date of this agreement and under the terms
  of this agreement, we commit to making available data obtained as
  part of the PTA projects. Access to earlier data will be by
  negotiation with the relevant group.
\item The status and terms of this agreement will be reviewed
  annually.
\end{enumerate}}

The IPTA {\bf Publication Policy} is another important document governing
the operation of the IPTA. It was ratified by the IPTASC in March 2012
and has the following items:

{\footnotesize
\begin{enumerate}[1]
\item Projects using or interpreting shared non-public IPTA data,
  hereafter ``IPTA projects'' must be approved by the IPTA Steering
  Committee and listed on the password-protected IPTA Projects website
  with a provisional title, abstract and list of principal authors for
  associated papers. IPTA data (hereafter ``data'') is any previously
  unreleased pulsar timing data from two or more PTA members of the
  IPTA.
\item Each PTA will recommend who among its members will be an author
  on an IPTA paper. Authorship order will be determined by the IPTASC
  considering any recommendation by the group of principal
  authors. There may be a small lead group of authors followed by a
  larger group of other authors listed alphabetically, or it may be a
  fully alphabetical list. Authorship will include at least everyone
  who contributed significantly to obtaining, processing, managing
  and/or interpreting the data discussed in the paper. Other
  contributions may also be worthy of authorship.
\item Any member of an IPTA PTA can request inclusion in the group of
  principal authors for a given project. Such requests will be decided
  by the IPTASC considering any recommendation by the existing group
  of principal authors.
\item Any member of the proposed author list for a paper may elect to
  opt out of authorship on that paper at any time prior to submission.
\item IPTA projects, including graduate student projects, will be
  protected from prior publication by others provided the project is
  completed in timely manner.
\item The status of all unpublished IPTA papers, including graduate
  student projects, will be reviewed at regular intervals by the
  IPTASC and, if deemed necessary by the SC, modifications made to the
  lead authorship group.
\item The IPTA encourages the early circulation of mature manuscript
  drafts to the entire author list. When the principal authors deem a
  manuscript ready for submission it will be circulated to all
  co-authors and the IPTASC for a review and comment period of at
  least two weeks.
\item Following the review and comment period, the revised manuscript,
  with a cover note addressing the comments received and their
  disposition, will be submitted to and reviewed by the
  IPTASC. Following IPTASC approval the corresponding author will
  simultaneously submit the final manuscript for publication and to
  all co-authors.
\item Upon receipt, all editorial correspondence (reviewer comments,
  acceptance/rejection notices, notices of on-line or other
  publication) related to a submitted IPTA manuscript will be
  circulated by the corresponding author to the IPTASC and to all
  co-authors. When the principal authors have prepared a response it
  will be circulated to the IPTASC and all co-authors. The IPTASC will
  decide whether the response requires a review and comment period
  and, if so, its duration. (Ordinarily, minor revisions will require
  no review and comment period, but major revisions will involve a
  review and comment period of at least one week.)
\end{enumerate}}

IPTA Projects operate under the terms contained in the Constitution
and the Publication Policy. Current projects that have been approved
by the IPTASC are listed in Table~\ref{tb:proj} along with the lead
proposer and the date of approval by the IPTASC. 

\begin{table}[ht]
{\footnotesize
\caption{IPTA Projects}\label{tb:proj}
\begin{tabular}{lll} \hline
\multicolumn{1}{c}{Project Title} & \multicolumn{1}{c}{Lead person} &
\multicolumn{1}{c}{Date Approved} \\ \hline
Pulsar Timescale & George Hobbs & 2012 June \\
Solar System Studies & Patrick Lazarus & 2012 September \\
IPTA Combined Data Release & Joris Verbiest & 2012 September \\
\hline
\end{tabular}}
\end{table}

The Pulsar Timescale project aims to use the IPTA combined data sets
to establish a realisation of terrestrial time, TT(IPTAxx) where 20xx
is the year of the realisation, that is of comparable or better
precision compared to the best long-term atomic timescales. This work is a
follow-up to the PPTA project that established TT(PPTA11) as described
by Hobbs et al. (2012)\nocite{hcm+12}. 

The Solar System Studies project builds on the work of Champion et
al. (2010)\nocite{chm+10} who used a subset of the PPTA data set plus
archival data from the Arecibo and Effelsberg telescopes to determine
the masses of solar system planetary systems. The best
result was obtained for the Jupiter system where the mass was obtained
with a precision of $2\times 10^{-10}$~M$_\odot$, comparable to the
best available measurements. In addition to improving planetary-system
masses, the IPTA project aims to search for solar-system bodies that
are not included in the solar-system ephemerides used for current
timing analyses and to constrain variations in the astronomical unit
over the past 20 years.

The remaining approved project is described in \S\ref{sec:data}
below. There are several proposed projects that are currently awaiting
discussion and approval by the IPTASC. 

\section{IPTA Data Sets}\label{sec:data}
Provision of combined data sets that contain well-calibrated results
from all three participating PTAs in a convenient form is fundamental
to the goals of the IPTA. This task is an IPTA project in its own
right (Table~\ref{tb:proj}) and all other IPTA projects depend on
it. Data sets from the different PTAs cover different pulsars,
different frequency bands, have different random and systematic noise
properties, different cadences and different data spans. Combining
these data sets in a consistent way with optimal weighting and
including corrections for dispersion variations is a non-trivial
problem and is a work-in-progress. Based on the EPTA, NANOGrav and
PPTA papers in this CQG Focus issue, Table~\ref{tb:data} lists the 50
pulsars for which data are currently (or soon will be) available on
the IPTA website. The pulsar name, period, dispersion measure and mean
flux density at 1400~MHz are given in the first four columns. The next
six columns give the data span in years and the number of ToA
epochs\footnote{Multi-channel ToAs from a single observation are
  counted as one ToA epoch.} for each pulsar observed by each of the
three main PTAs. The final column gives the total span of the combined
data sets expressed as a range of Modified Julian Dates.

\begin{table}[ht]
{\scriptsize
\caption{Parameters of the PTA and IPTA Data Sets \\}\label{tb:data}
\begin{tabular}{lccc|cc|cc|cc|c} \hline
& & & & \multicolumn{2}{c|}{EPTA}&\multicolumn{2}{c|}{NANOGrav}
&\multicolumn{2}{c|}{PPTA} & IPTA \\
\multicolumn{1}{c}{Pulsar}& P$_0$ & DM & S$_{1400}$ & Span & Nr
& Span & Nr & Span & Nr & Span \\
& (ms) & (cm$^{-3}$pc) & (mJy) & (yr) & ToAs & (yr) & ToAs & (yr) 
& ToAs &(MJD) \\ \hline 
&&&&&&&&&&\\
J0023+0923   &  3.05 &  14.3 &  ...& ... & ... &  1.9& 56  &...  &...  &55730--56438\\
J0030+0451   &  4.87 &   4.3 &  0.6& 12.8&  737& 15.5& 324 &...  &...  &50787--56438\\
J0340+4130   &  3.29 &  49.6 &  ...& ... & ... &  1.3& 41  &...  &...  &55971--56432\\
J0437$-$4715 &  5.76 &   2.6 &149.0&...  & ... & ... &     & 17.0& 5160&50190--56379\\
J0613$-$0200 &  3.06 &  38.8 &  2.3&14.2 & 1192&  8.4& 217 & 13.3& 799 &50931--56432\\ 
&&&&&&&&&&\\
J0645+5158   &  8.85 &  18.2 &  ...& ... & ... &  2.0&  53 &...  &...  &55699--56432\\
J0711$-$6830 &  5.49 &  18.4 &  3.2&...  & ... &...  &...  & 19.2& 729 &49373--56396\\
J0751+1807   &  3.48 &  30.2 &  3.2&15.3 & 3199& 5.9 &  68 &...  &...  &50363--55948\\
J0931$-$1902 &  4.64 &  41.5 &  ...& ... & ... & 0.2 &  8  &...  &...  &56350--56432\\
J1012+5307   &  5.26 &   9.0 &  3.0& 15.2& 1195& 8.8 & 233 &...  &...  &50362--56432\\ 
&&&&&&&&&&\\
J1017$-$7156 &  2.34 &  94.2 &  0.9&...  & ... &...  &...  & 2.6 & 248 &55456--56395\\
J1022+1001   & 16.45 &  10.2 &  6.1& 15.2& 813 &15.5 & 107 & 10.3& 755 &50361--56441\\
J1024$-$0719 &  5.16 &   6.5 &  1.5& 15.0&  432& 3.7 & 120 & 17.2& 626 &50117--56432\\
J1045$-$4509 &  7.47 &  58.2 &  2.7&...  & ... &...  &...  & 19.1& 714 &49405--56395\\
J1455$-$3330 &  7.99 &  13.6 &  1.2&...  & ... & 8.8 & 232 &...  &...  &53216--56432\\ 
&&&&&&&&&&\\
J1600$-$3053 &  3.60 &  52.3 &  2.5& 5.3 &  380& 5.6 & 173 & 11.2& 754 &52031--56432\\
J1603$-$7202 & 14.84 &  38.0 &  3.1&...  & ... &...  &...  & 17.4& 590 &50026--56395\\
J1614$-$2230 &  3.15 &  34.5 &  ...& ... & ... &  4.7& 134 &...  &...  &54723--56432\\
J1640+2224   &  3.16 &  18.4 &  2.0& 15.0&  517& 15.5& 279 &...  &...  &50459--56438\\
J1643$-$1224 &  4.62 &  62.4 &  4.8& 14.9& 659 & 8.8 & 236 & 19.1& 550 &49421--56432\\ 
&&&&&&&&&&\\
J1713+0747   &  4.57 &  16.0 & 10.2& 18.4& 1144& 21.2& 857 & 19.1& 683 &48737--56466\\
J1730$-$2304 &  8.12 &   9.6 &  3.9& 14.2& 191 & 4.2 & 21  & 19.1& 514 &49421--56395\\
J1732$-$5049 &  5.31 &  56.8 &  1.7&...  & ... &...  &...  &  8.4& 226 &52647--55724\\
J1738+0333   &  5.85 &  33.8 &  ...& 5.0 &  211& 3.9 & 79  &...  &...  &54103--56426\\
J1741+1351   &  3.75 &  24.0 &  0.9& ... & ... & 4.0 & 100 &...  &...  &54997--56446\\ 
&&&&&&&&&&\\
J1744$-$1134 &  4.07 &   3.1 &  3.1& 15.0&  394& 8.8 & 226 & 18.2& 604 &49729--56432\\
J1747$-$4036 &  1.65 & 152.9 &  ...& ... & ... & 1.2 &  41 &...  &...  &55975--56432\\
J1824$-$2452A&  3.05 & 120.5 &  2.0&...  & ... &...  &...  & 7.9 & 398 &53518--56394\\
J1853+1303   &  4.09 &  30.6 &  0.4& 5.9 & 75  & 8.4 & 126 &...  &...  &53369--56438\\
J1857+0943   &  5.36 &  13.3 &  5.0& 14.9&  389& 27.4& 853 & 9.1 & 416 &46435--56438\\ 
&&&&&&&&&&\\
J1903+0327   &  2.15 & 297.5 &  1.3& ... & ... & 5.7 &  82 &...  &...  &54356--56426\\
J1909$-$3744 &  2.95 &  10.4 &  2.1&  7.0&  226& 8.8 & 212 &10.3 & 1189&52618--56432\\
J1910+1256   &  4.98 &  38.1 &  0.5& ... & ... & 8.4 & 127 &...  &...  &53369--56438\\
J1911+1347   &  4.63 &  31.0 &  0.1& 5.0 &  90 &...  & ... &...  &     &54092--56341\\
J1918$-$0642 &  7.65 &  26.6 &  0.6&10.5 & 216 & 8.8 & 218 &...  &...  &52095--56440\\ 
&&&&&&&&&&\\
J1923+2515   &  3.79 &  18.9 &  ...& ... & ... & 2.6 & 62  &...  &...  &55492--56446\\
J1939+2134   &  1.56 &  71.0 & 13.2&15.3 & 1109& 28.6&1222 & 17.6& 489 &45984--56446\\
J1944+0907   &  5.19 &  24.3 &  ...& ... & ... & 5.3 & 103 &...  &...  &54504--56426\\
J1949+3106   & 13.14 & 164.1 &  0.2& ... & ... & 0.8 &  26 &...  &...  &56138--56425\\
J1955+2908   &  6.13 & 104.5 &  1.1&...  & ... & 28.2& 144 &...  &...  &46111--56426\\ 
&&&&&&&&&&\\
J2010$-$1323 &  5.22 &  22.2 &  1.6& 5.1 &  302& 4.7 & 125 &...  &...  &54086--56432\\
J2017+0603   &  2.90 &  23.9 &  0.5& ... & ... & 2.6 & 69  &...  &...  &55499--56438\\
J2043+1711   &  2.38 &  20.7 &  ...& ... & ... & 1.9 & 85  &...  &...  &55730--56426\\
J2124$-$3358 &  4.93 &   4.6 &  3.6& ... & ... & 4.3 &  17 & 18.9& 713 &49489--56394\\
J2129$-$5721 &  3.73 &  31.9 &  1.1&...  & ... &...  &...  & 17.5& 576 &49987--56396\\ 
&&&&&&&&&&\\
J2145$-$0750 & 16.05 &   9.0 &  8.9& 15.2&  693& 8.8 & 187 & 18.8& 751 &49517--56432\\
J2214+3000   &  3.12 &  22.6 &  ...& ... & ... & 3.6 &  65 &...  &...  &55121--56438\\
J2241$-$5236 &  2.19 &  11.4 &  4.1&...  & ... &...  &...  & 3.2 & 247 &55235--56394\\
J2302+4442   &  5.19 &  13.8 &  1.2& ... & ... & 1.3 & 39  &...  &...  &55971--56432\\
J2317+1439   &  3.45 &  21.9 &  4.0& 14.9& 406 & 20.7& 406 &...  &...  &48861--56438\\
\hline
\end{tabular}}
\end{table}

\begin{figure}[ht]
\centerline{\includegraphics[angle=-90,width=160mm]{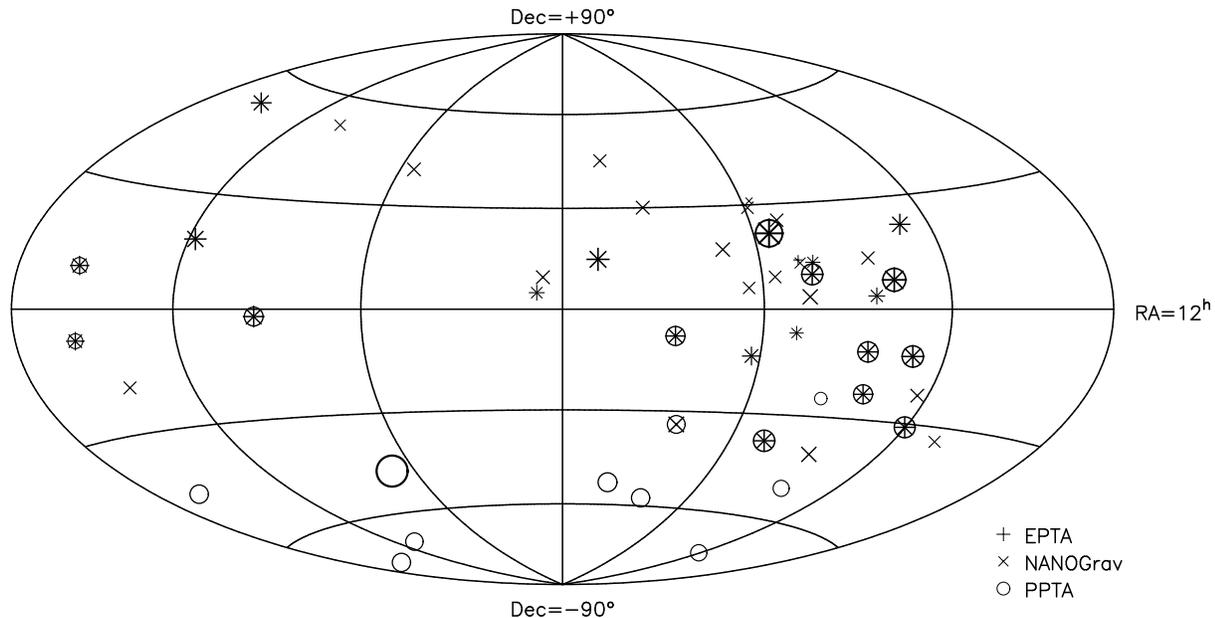}}
\caption{Distribution on the sky of the pulsars being observed by
  participating PTAs. Right ascension increases to the left and
  $0^{\rm h}$ is at the centre of the plot. The
  symbol sizes are larger for larger values of the ratio
  S$_{1400}$/P$_0$. (Pulsars without published values of S$_{1400}$ are
  assumed to have S$_{1400}=1.0$~mJy.)}\label{fg:ipta_sky}
\end{figure}

\begin{figure}[ht]
\centerline{\includegraphics[angle=-90,width=160mm]{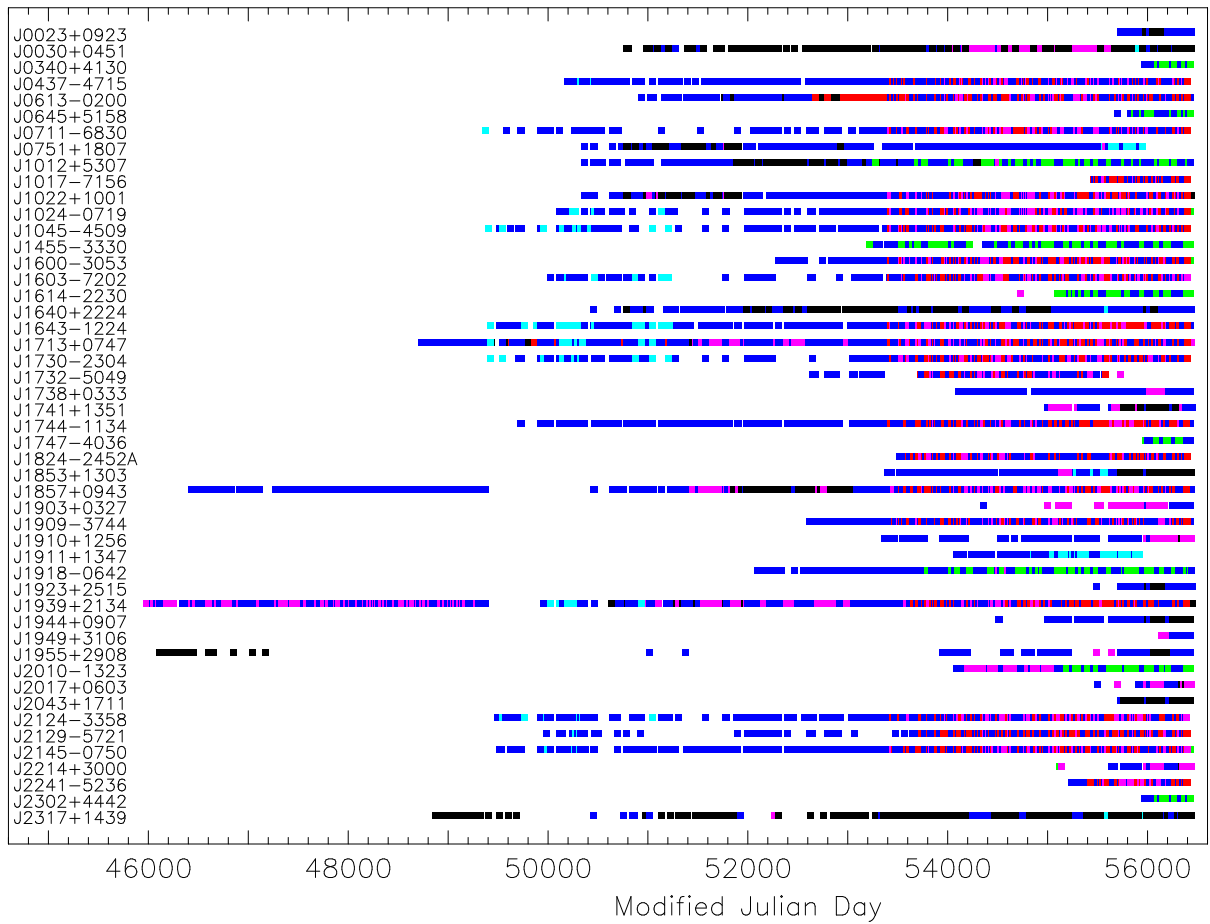}}
\caption{Distribution of ToAs for the combined data sets. The band centre
  frequencies ($\nu$) are colour encoded as follows: black:
  $\nu<$~500~MHz, red: 500~MHz~$<\nu<$~750~MHz, green:
  750~MHz~$<\nu<$~1000~MHz, blue: 1000~MHz~$<\nu<$~1500~MHz, aqua:
  1500~MHz~$<\nu<$~2000~MHz and pink:
  2000~MHz~$<\nu<$~4000~MHz.}\label{fg:ipta_spans}
\end{figure}

Figure~\ref{fg:ipta_sky} shows the distribution on the sky of pulsars
being observed by the PTAs participating in the IPTA and
Figure~\ref{fg:ipta_spans} shows the time coverage of ToAs for the
combined data sets. Not unexpectedly, the sky distribution is somewhat
uneven with fewer observed pulsars in the northern hemisphere and a
concentration in the central regions of the Galaxy around 18$^{\rm h}$
right ascension where the density of pulsars is greater. The optimal
sky coverage is a complex issue dependent on the application. For
example, for the Pulsar Timescale project it is irrelevant, whereas for
detection of an isotropic background of GW a wide range of angular
separations on the sky is required but sky position itself is not
important. For detection of an isolated source of GW, the optimal
distribution depends on the direction of the source (see, e.g., Boyle
\& Pen, 2012).\nocite{bp12} Of the 50 IPTA pulsars, 12 are being
observed by all three PTAs and 10 by two PTAs. While some overlap is
desirable, in fact essential, the current degree of overlap seems
somewhat excessive. Scheduling of PTAs with multiple telescopes while
ensuring adequate observation cadence and frequency coverage is not
trivial \cite{lbj+12}, but it is likely that addressing the issue
would result in some improvement in the overall efficiency and
effectiveness of the IPTA.

Four IPTA pulsars have data spans longer than 20 years: PSR J1713+0747
(21.2 years), PSR J1857+0943 (PSR B1855+09, 27.4 years), PSR
J1939+2134 (PSR B1937+21, 28.6 years) and PSR J1955+2908 (PSR
B1953+29, 28.2 years). The early observations for all four of these
pulsars were made at Arecibo \cite{fwc93,ktr94,rtd88}. Unfortunately,
except for J1713+0747, there are substantial gaps between the early
and later data, and so arbitrary jumps are generally required between
the data sets and pulse counting ambiguities are possible,
signficantly reducing the value of the early data. Since 2005
(MJD $\sim 53400$) the cadence of observations has improved
dramatically and there is more overlap of data at different frequency
bands or with different instruments. However, the long data spans are
important for detection of the red signals expected from GW and other
sources despite the reduced cadence. For three pulsars: J0751+1807,
J1732$-$5049 and J1911+1347, data-taking appears to have ceased a year
or two ago. The PPTA dropped J1732$-$5049 since its ToAs had much
larger uncertainties than those of other nearby pulsars.

\section{Meetings and Outreach}\label{sec:outreach}
One of the important roles of the IPTA is to facilitate communication
between the members of the participating PTAs, with the broader
scientific community and with the general public. Aside from the wiki
pages described in \S\ref{sec:org}, the main avenue for communication
between PTA members is the series of now-annual science meetings and
the associated student workshops. Since 2010, these
meetings have been generously supported by the National Science
Foundation's ``Partnerships for International Research and Education''
(PIRE) grant, awarded to West Virginia University and NANOGrav in 2010
with Maura McLaughlin as Principal Investigator. 

The first ``IPTA'' meeting was held in Arecibo on 1-2 August,
2008. The meeting was attended by about 35 people representing all
three PTAs and the larger gravitational-wave community and including 7
or 8 students. It included a celebratory dinner in honour of Don
Backer's 65th birthday. There were four sessions on, respectively, the
status of the three PTAs, techniques to improve timing precision,
connections between the laser-interferometer projects and PTAs, and
the future development of PTA projects. All sessions had extensive
discussion periods during which many of the ideas that were later
realised with the formation of the IPTA were discussed.

The second IPTA
meeting\footnote{http://www.lorentzcenter.nl/lc/web/2010/388/info.php3?wsid=388\&venue=Oort}
was held at the Lorentz Center in Leiden between 21 June and 2 July
2010, supported by the NSF PIRE grant. The meeting was attended
by about 65 people, nearly half of whom were students. This was the
first IPTA meeting to have a pedagogical Student Workshop; this
extended over five days (21 -- 25 June) and included lectures by
experts and practical tutorial sessions on gravitational-wave basics,
pulsar timing basics, the effects of the interstellar medium, and data
analysis techniques for GW detection (two days). The Science Meeting
was held on the following week and followed a similar set of session
headings on each of the first four days with both invited and
contributed talks and a Discussion session at the end of each day. On
the fifth day, there were presentations on each of the three main PTAs
followed by discussion sessions to end the meeting.

The 2011 meeting\footnote{http://ipta.phys.wvu.edu/ipta-2011/} was
held in West Virginia and organised by NANOGrav. The meeting was also
notable for the large number of students attending; of the 86 people
attending the Student Workshop and/or the Science meeting, more than
half were students, with 27 graduate students, 20 undergraduate
students and one high school student. The Student Workshop was held on
6 -- 10 June at West Virginia University, with Xavier Siemens in
charge of its organisation. The format of the Workshop was very
similar to the Leiden meeting with introductory lectures followed by
hands-on tutorial sessions on each of the days. The Science Meeting
was held at Snowshow Resort in the Allegheny Mountains, just 7.5 km
line-of-sight from Green Bank, on 13 -- 17 June. There were more than
25 invited talks and extensive discussions each day. A significant
part of the discussion concerned the organisation and operation of the
IPTA, especially the project and publication guidelines, and consensus
was largely reached on the important issues.

The 2012 meeting\footnote{http://ipta.phys.wvu.edu/ipta-2012/} was
hosted by the PPTA. The Student Workshop was held at
Sydney University on 18 -- 22 June and the Science Meeting was held at
Kiama, a seaside town about 100 km south of Sydney, on 25 -- 29 June. A
total of 84 people, including 29 students and eight participants from
China, attended the meetings with most attending both the workshop and
the science meeting. The Workshop followed a similar format to
previous meetings with highlights being a session of live observing
on the Parkes 64-m radio telescope using the Pulse@Parkes system
\cite{hhc+09} and a talk on aboriginal astronomy by Ray Norris. The
Science Meeting began with reviews of the current status of the three
main PTAs and then presentations and discussions on data analysis,
including several talks on cyclic spectroscopy, GW detection algorithms,
implications of limits on GW signals, other applications of PTA data
sets, optimisation of PTA/IPTA observing strategies and future
telescopes and instrumentation, including a presentation by Di Li on
the Chinese FAST project. 

The 2013 IPTA meetings\footnote{http://ipta.phys.wvu.edu/} were held
17 -- 28 June in Krabi, Thailand, approximately equi-distant from all
participating PTA centres, with Busaba Kramer playing
a major role in the local organising. About 90 people attended the
meeting, including more than 30 students. These included many who had
not previously attended an IPTA meeting. The Student Workshop was very
successful, concentrating on signal processing for GW detection, and
was a little more relaxed than previous years with more time for
informal discussions (and relaxing by the pool!). Highlights of the
Science Meeting included the first analyses of real IPTA data sets and
extensive discussions on data analysis methods, especially considering
the effects of red and white timing noise. Presentations and
discussions on the nature of the nanohertz GW background, especially
its likely anisotropy, and projections of GW sensitivity for future
IPTA data sets were also of great interest.

The terms of the PIRE grant require that the IPTA meetings be formally
evaluated; this is achieved by means of questionnaires distributed at
the meeting. Summaries of these evaluations may be found in the
Quarterly Newsletters on the PIRE
website.\footnote{http://nanograv-pire.wvu.edu/pire.html\#evaulation}

Plans are already coming together for the 2014 IPTA meetings, to be
organised by NANOGrav and held at the University of Calgary (Workshop)
and Banff National Park (Science Meeting). We plan to have the 2015
meetings in South Africa, to be organised by the PPTA in collaboration
with Sarah Buchner, probably Durban for the Workshop and Kruger
National Park for the Science Meeting. Note that participation in the
IPTA Workshops and Meetings is not restricted to members of the
participating PTAs and attendance by others from the larger astronomy
and astrophysics community is encouraged.

The IPTA has a Working Group led by Ryan Lynch that focusses on
outreach to the general public. The internal web page, accessible to
group members, lists the outreach activities of the three PTAs and also
acts as a forum for discussion of ideas related to IPTA outreach. The
public website\footnote{http://www.ipta4gw.org} also has an outreach
page giving links to PTA outreach sites. 

Detection of GW signals in pulsar timing data is a complex task, with
many possible approaches. To provide some confidence in the techniques
being used by various groups around the world, not all associated with
one of the three main PTAs, the IPTA has hosted a Data Challenge with
a small Working Group led by Rick Jenet. This Data Challenge has made
several simulated PTA data sets publically available as part of Data
Challenge 1 and invited interested groups to analyse them. There are
two classes of data set, Open and Closed, both of which are somewhat
idealised and lacking all the complications of real data
sets. Parameters of the injected GW signal were provided for the Open
data sets, but were hidden for the Closed sets. A total of 12 analyses
of the Closed data sets were submitted by the deadline of 28
September, 2012, most of which obtained results consistent with
expectations.\footnote{See the Data Challenge pages at
  http://www.ipta4gw.org and the article by Rick Jenet in this Special
  Issue.} Further, more realistic, data challenges are planned for the
future.

Over the past few years, pulsar timing has been increasingly
recognised as an important tool in the quest to detect GW,
complementary to the various laser-interfometer systems, existing and
planned. Stan Whitcomb, as Executive Secretary of the Gravitational
Wave International Committee (GWIC), attended the 2008 Arecibo IPTA
meeting, and promoted the idea of the PTAs joining GWIC. Shortly after
this meeting, GWIC invited representatives of the three PTAs to join
the 21 representatives of 11 GW detector projects and related bodies
on the Committee. This invitation was accepted on 22 November, 2008,
with Michael Kramer, Andrea Lommen and Dick Manchester representing
the EPTA, NANOGrav and the PPTA respectively. The current PTA
representatives are Michael Kramer, Rick Jenet and George Hobbs. Among
other activities, GWIC organises an annual GWIC Thesis Prize,
typically with 20 or so theses submitted each year. The 2011 prize was
awarded to Rutger van Haasteren, a member of the EPTA, for his thesis
``Gravitational Wave Detection and Data Analysis for Pulsar Timing
Arrays'', a notable first for the PTA community.

\section*{Acknowledgements}
This paper describes work done by the many people who
participate in the activities of the IPTA and I thank them for their
efforts in support of the organisation.

\section*{References}


\end{document}